\pdfoutput=1
\documentclass[manuscript,screen,authordraft=false,review=false,timestamp=false,nonacm]{acmart}

\usepackage{algorithmic}
\usepackage{graphicx}
\usepackage{textcomp}
\usepackage{xcolor}
\usepackage{subfigure}
\usepackage{listings}
\usepackage{hyperref}

\AtBeginDocument{%
  }

\setcopyright{acmlicensed}
\copyrightyear{2024}
\acmYear{2024}
\acmDOI{XXXXXXX.XXXXXXX}

\makeatletter
\def\lst@makecaption{%
  \def\@captype{table}%
  \@makecaption
}
\makeatother
    
\begin{document}

\title{Revisiting VerilogEval: A Year of Improvements in Large-Language Models for Hardware Code Generation}

\author{Nathaniel Pinckney}
\email{npinckney@nvidia.com}
\orcid{0000-0001-6159-8964}
\affiliation{%
  \institution{NVIDIA Corporation}
  \city{Austin}
  \state{Texas}
  \country{USA}
}

\author{Christopher Batten}
\email{cbatten@cornell.edu}
\orcid{0000-0002-2835-667X}
\affiliation{%
  \institution{NVIDIA Corporation}
  \city{Santa Clara}
  \state{California}
  \country{USA}
}
\affiliation{%
  \institution{Cornell University}
  \city{Ithaca}
  \state{New York}
  \country{USA}
}

\author{Mingjie Liu}
\email{mingjiel@nvidia.com}
\orcid{0000-0002-3488-9763}
\affiliation{%
  \institution{NVIDIA Corporation}
  \city{Austin}
  \state{Texas}
  \country{USA}
}

\author{Haoxing Ren}
\email{haoxingr@nvidia.com}
\orcid{0000-0003-1028-3860}
\affiliation{%
  \institution{NVIDIA Corporation}
  \city{Austin}
  \state{Texas}
  \country{USA}
}

\author{Brucek Khailany}
\email{bkhailany@nvidia.com}
\orcid{0000-0002-7584-3489}
\affiliation{%
  \institution{NVIDIA Corporation}
  \city{Austin}
  \state{Texas}
  \country{USA}
}




\begin{abstract}
The application of large-language models (LLMs) to digital hardware code generation is an emerging field, with most LLMs primarily trained on natural language and software code. Hardware code like Verilog constitutes a small portion of training data, and few hardware benchmarks exist. The open-source VerilogEval benchmark, released in November 2023, provided a consistent evaluation framework for LLMs on code completion tasks. Since then, both commercial and open models have seen significant development.

In this work, we evaluate new commercial and open models since VerilogEval's original release—including GPT-4o, GPT-4 Turbo, Llama3.1 (8B/70B/405B), Llama3 70B, Mistral Large, DeepSeek Coder (33B and 6.7B), CodeGemma 7B, and RTL-Coder—against an improved VerilogEval benchmark suite. We find measurable improvements in state-of-the-art models: GPT-4o achieves a 63\% pass rate on specification-to-RTL tasks. The recently released and open Llama3.1 405B achieves a 58\% pass rate, almost matching GPT-4o, while the smaller domain-specific RTL-Coder 6.7B models achieve an impressive 34\% pass rate. 

Additionally, we enhance VerilogEval’s infrastructure by automatically classifying failures, introducing in-context learning support, and extending the tasks to specification-to-RTL translation. We find that prompt engineering remains crucial for achieving good pass rates and varies widely with model and task. A benchmark infrastructure that allows for prompt engineering and failure analysis is essential for continued model development and deployment.
\end{abstract}

\ccsdesc[500]{Hardware~Hardware description languages and compilation}
\ccsdesc[500]{Computing methodologies~Machine learning}

\keywords{large language models, RTL code generation, benchmarks}


\maketitle

\section{Introduction}


Applications of large-language models (LLMs) to software coding have reached wide deployment, with examples such as GitHub Copilot \cite{copilot_github_2024}. Yet, applications of LLMs to hardware design are still in their infancy \cite{Blocklove_2023,chang-chipgpt-arxiv2023}. Hardware code generation benchmarks have only been available since 2023, including RTLLM \cite{lu2023rtllm}, VerilogEval \cite{liu2023verilogeval}, VeriGen~\cite{thakur-verilog-codegen-date2023,thakur-verigen-todaes2024}, and most recently RTL-Repo \cite{allam2024rtlrepo}. Despite this, LLM model releases have been extremely rapid. In this work, we survey the progress LLMs have been in the past year by evaluating newer large-language models than those tested in the original VerilogEval paper (published November 2023), including GPT-4o \cite{gpt4o} and GPT-4 Turbo \cite{gpt4_turbo_announce}, open-source models like Llama3.1 \cite{noauthor_meta-llamallama3_2024}, and domain-specific models such as RTL-Coder \cite{liu2024rtlcoder}. In short, we assess the latest state-of-the-art language models to determine the current frontier of LLM-based Verilog code generation while also evaluating the impact of prompt tuning. We find that recent open models are competitive with closed models, and that prompt tuning varies considerably across models.

We also take the opportunity to release an improved version of VerilogEval to better align with instruction-tuned models and to encourage further prompt tuning research. While RTLLM benchmarked  conversational specification-to-RTL generation performance, VerilogEval, VeriGen, RTL-Repo are code completion benchmarks. Additionally, none of the benchmarks explore a model's generation performance using in-context learning \cite{brown2020language} examples, nor do they provide a detailed way to inspect the reasons for a model's failure.

This work aims to address these limitations by extending VerilogEval \cite{liu2023verilogeval} (henceforth known as ``VerilogEval v1'') to support specification-to-RTL tasks in addition to the original code completion task.  We also incorporate a variable number of in-context learning prompts, and provide a robust failure classification mechanism, to provide a more comprehensive evaluation framework for Verilog code generation tasks. The significance of these improvements is its potential to push LLM development forward for hardware design, through offering insights into model performance and the efficacy of prompt tuning, and to point out differences in generation quality across tasks. Even with similar problem statements and in-context learning examples, we find divergent responses by large-language models. This variability highlights the importance of understanding how different models respond to various prompts and contexts through the use of the benchmarks, providing granular failure feedback.


The following new features are part of the improved ``VerilogEval v2'' benchmark infrastructure:

\begin{enumerate}
    \item \textit{Specification-to-RTL task support}: VerilogEval v1 only supported code completion tasks, such as used in Copilot\cite{copilot_github_2024}, while many models are tuned and deployed as instruction-tuned models\cite{yuan2023evaluating}, with question and answer prompting. 

    \item \textit{In-context learning examples}: No in-context learning (ICL) \cite{brown2020language} examples were supported as part of the prompt in VerilogEval v1. Prompt tuning techniques, such as in-context learning, can improve LLM responses.
    
    \item \textit{Failure classification}: VerilogEval v1 only reported pass/fail results of a benchmark problem, and did not give fine-grained feedback on failures.

   \item \textit{Makefile-based evaluation environment}: The original VerilogEval benchmark \cite{liu2023verilogeval} used a monolithic dataset, whereas the proposed infrastructure uses a Makefile-based approach. This allows for easier scaling while sweeping evaluation settings across more models than the original benchmark, and easier human inspection of the dataset.
\end{enumerate}

The improved VerilogEval benchmark is available publicly at \url{https://github.com/NVlabs/verilog-eval}.
\vspace{3mm}

\section{VerilogEval v1 Revisited}

The original VerilogEval \cite{liu2023verilogeval} contains 156 problems adopted from questions and solutions on the HDLBits Verilog instructional website, picked based on clarity and diversity. The VerilogEval dataset contains both VerilogEval-machine and VerilogEval-human problem descriptions. The former is based on GPT-3.5 generated descriptions of the solution RTL, while the latter are human-created problem descriptions from HDLBits. While both problem description sets are useful when evaluating a large language model's code generation capability, VerilogEval-human is most aligned with common code generation deployments, such as Copilot \cite{copilot_github_2024}. 

VerilogEval evaluated an LLM's performance by reporting pass rate, specifically a $pass@k$ metric meaning at least 1 sample passes among $k$ samples. Because LLM responses will be non-deterministic at non-zero temperature, the LLM is sampled multiple times and, if at least one sample passes, the problem is passed. If none of the $k$ samples pass, then the problem is failed. Formally $pass@k$ is defined as follows:
\begin{equation}
    pass@k := \mathbb{E}_{\text{Problems}} \left[ 1 - \frac{\binom{n-c}{k}}{\binom{n}{k}} \right],
\end{equation}
where $n \geq k$ represents the total number of trials for each problem, and $c$ represents the number of trials that pass the functional check. 

The original VerilogEval paper reported results with $k = 1, 5, 10$. This $pass@k$ metric is useful to evaluate if given knowledge is available within an LLM, and thus beneficial for model development, but a typical LLM deployment for code generation (say, in an interactive copilot application \cite{copilot_github_2024} will only sample for a response once). Therefore, for the evaluation of LLMs in this work, only $pass@1$ is reported to mimic single turn (single query and response) scenarios. However, we report two sets of model parameters: high temperature (T=0.8, top\_p=0.95) and low temperature (T=0.0, top\_p=0.01) set. The high-temperature model parameters are identical to those used in \cite{liu2023verilogeval}. For high temperature, we report $pass@1$ across 20 samples (n=20). In other words, we report how many of the 20 sampled responses pass the benchmark for each problem within the dataset. For low temperature (nearly equivalent to greedy sampling), where responses are generally deterministic, we report a single sample (n=1).

For this study, we only evaluated models against VerilogEval-human to highlight the most useful LLM evaluation results. VerilogEval-machine in comparison, can be overly descriptive compared to real-world code generation problems. The infrastructure was revised to more easily re-run a subset of the dataset (discussed in the next session), to better post-process LLM response, and 14 problems had their descriptions or test benches revised to fix consistency or clarity issues. Beyond these changes, and minors white space differences, the dataset, and prompts were kept design same as the study in \cite{liu2023verilogeval}. In \cite{liu2023verilogeval} the highest achieving model was GPT-4 \cite{gpt4} with a $pass@1$ pass rate of $43.5\%$ and $60.0\%$ for VerilogEval-human and VerilogEval-machine, respectively. GPT-3.5 exhibited lower pass rates of $26.7\%$ and $46.7\%$.

We evaluate fourteen publicly available large-language models against these VerilogEval-human code completion prompts: 

\begin{itemize}
\item OpenAI GPT-4o \cite{gpt4o}
\item OpenAI GPT-4 Turbo (gpt-4-1106-preview) \cite{gpt4_turbo_announce}
\item OpenAI GPT-4 (gpt-4-0613) \cite{gpt4}
\item Mistral AI Mistral Large \cite{ai_au_2024}
\item Meta Llama3.1 405B, 70B, and 8B \cite{noauthor_meta-llamallama3_2024}
\item Meta Llama3 70B \cite{noauthor_meta-llamallama3_2024}
\item Meta Llama2 70B \cite{noauthor_meta-llamallama3_2024}
\item Meta CodeLlama 70B \cite{noauthor_meta-llamacodellama-70b-instruct-hf_2024}
\item Google CodeGemma 7B \cite{noauthor_googlecodegemma-7b_2024}
\item DeepSeek Coder 33B and 6.7B \cite{guo2024deepseekcoder} 
\item RTL-Coder DeepSeek v1.1 6.7B \cite{liu2024rtlcoder}
\end{itemize}

\begin{figure}[htbp]
    \centering
    \includegraphics[trim=10cm 3cm 10cm 3cm,clip,scale=0.5]{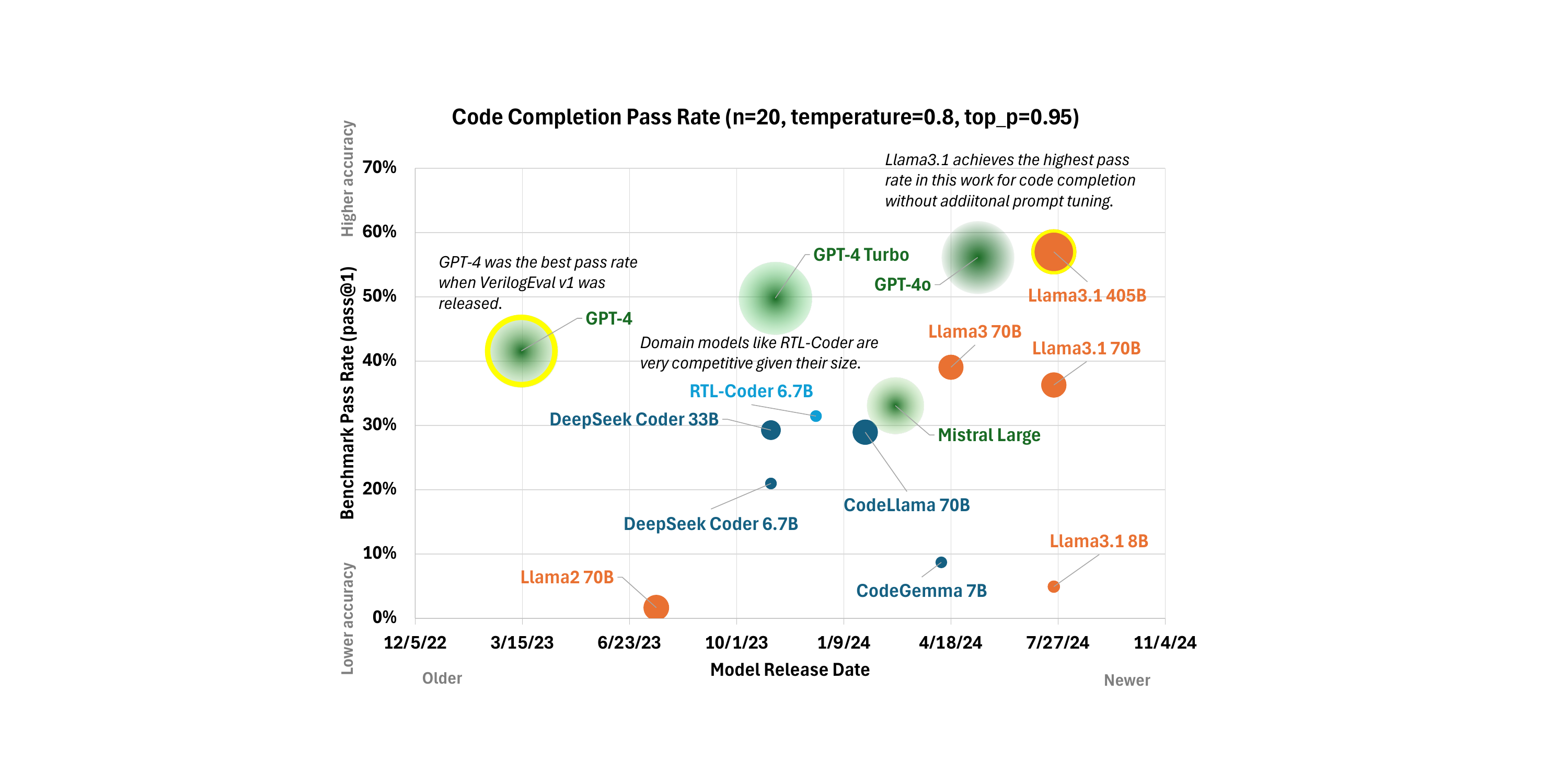}
    \caption{Pass rate across recent large-language models similar to VerilogEval v1 for pass@1. Green models are closed general-purpose models, orange are open general-purpose models, dark blue are coding-specific models, and light blue is an RTL-specific model.}
    \label{fig:survey}
\end{figure}

The models are comprised of a range of closed and open source, parameter sizes, and general-purpose to specialized. Figure \ref{fig:survey} shows the equivalent of VerilogEval-human pass@1 results for recent models, with pass rate on the y-axis and model release date on the x-axis. The data point size represents the approximately model size. Note that GPT-4, GPT-4 Turbo, GPT-4o, and Mistral Large have undisclosed sizes. The data points are green in colored for undisclosed size, orange for general-purpose open models, dark blue for coding-specific models, and light blue for domain-specific (RTL code generation) models.

GPT-4 with our new infrastructure and adjusted prompts is slightly lower than previously measured in \cite{liu2023verilogeval}, at $41.6\%$ instead of $43.5\%$.  This can be attributed mostly to slight changes in prompt, such as whitespace (line breaks) and punctuation, and is in the measurement noise. Of the large models, GPT-4o (56.1\%) and GPT-4 Turbo ($49.8\%$) exceed GPT-4, while Llama3.1 405B goes further still with the best pass rate ($57.0\%$) despite being an open model. Coding specific models such as DeepSeek Coder 33B ($29.3\%$) and CodeLlama 70B ($29.0\%$) did well for their size at their respective times of release. Significant improvement is seen from Llama2 70B ($1.7\%$) to Llama3 70B ($39.1\%$), while Llama3.1 70B (35.3\%) is slightly worse than its predecessor. However, as we shall see in the next section, prompt tuning can change the pass rate, significantly in some cases. Mistral Large ($33.1\%$) and CodeGemma 7B ($8.7\%$) appear to lag Verilog code generation tasks compared to their peers. Notably, RTL-Coder 6.7B ($31.5\%$) performs almost as well as Llama3.1 70B ($36.3\%$) while being an order of magnitude smaller, demonstrating the efficiency of smaller and cheaper domains-specific models. Overall, LLMs have demonstrated tremendous progress across model releases, especially open and domain-specific models.

An LLM's quality of result is not solely dependent on the model; prompt tuning can have a large impact on generated code quality. In the next section, we extend our new VerilogEval v2 benchmark to support two prompt tuning techniques: 1) in-context learning \cite{brown2020language} and 2) changing the task type from code completion to specification-to-RTL. In-context learning has been demonstrated to have similar impact to training or tuning \cite{dai2023gptlearnincontextlanguage,akyürek2023learningalgorithmincontextlearning,vonoswald2023transformerslearnincontextgradient,hendel2023incontextlearningcreatestask} while many models are instruction-tuned, such that specification-to-RTL may be better aligned with how models are created.

\vspace{3mm}

\section{VerilogEval v2 Improvements}

\begin{figure*}[hbtp]
    \centering
    \includegraphics[trim=0cm 0cm 0cm 0cm,clip,scale=0.3]{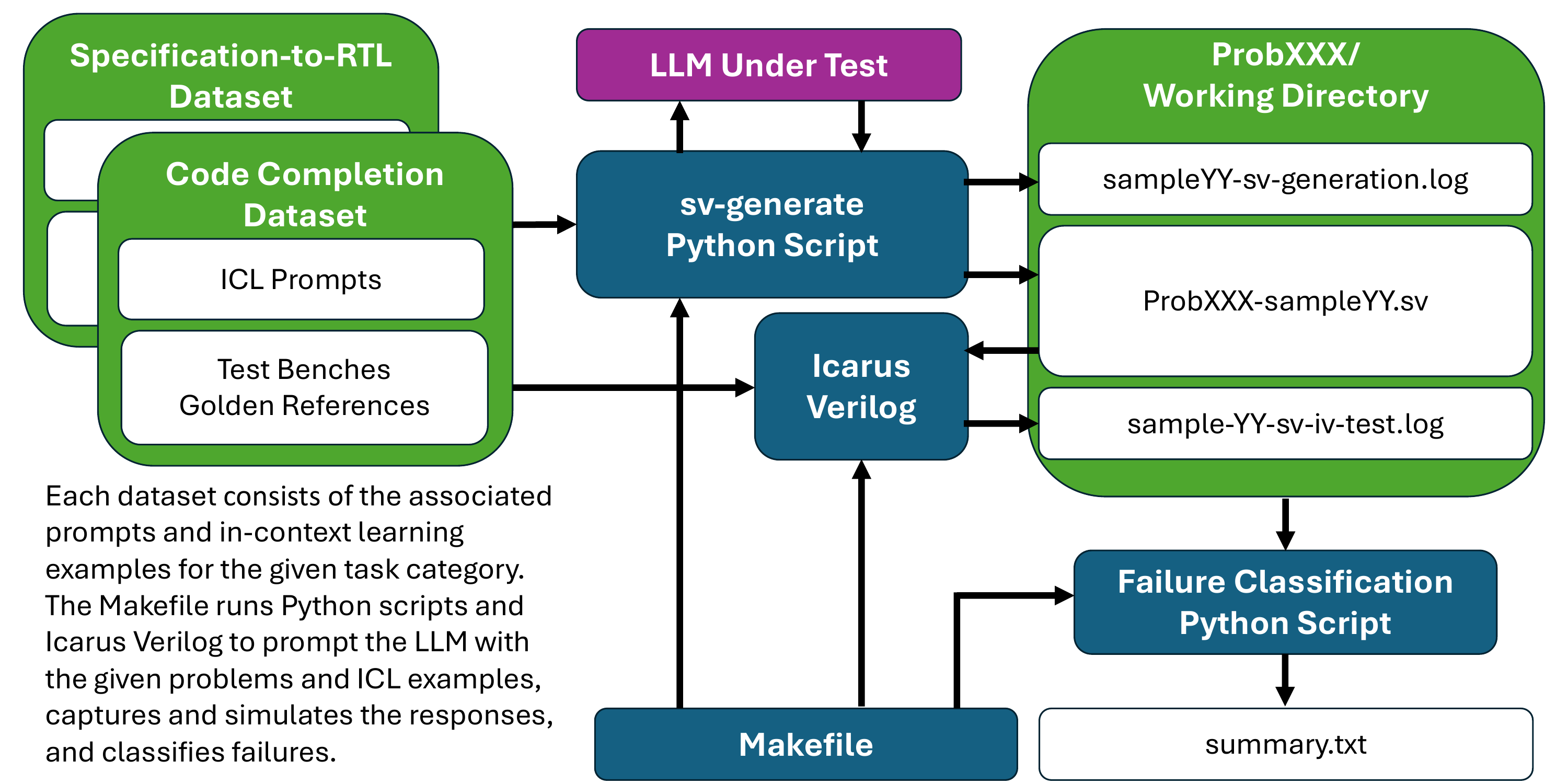}
    \caption{Overview of VerilogEval v2 flow.}
    \label{fig:flow_overview}
\end{figure*}

The improved VerilogEval v2 flow is shown in Figure~\ref{fig:flow_overview}, which is substantially different from the flow from \cite{liu2023verilogeval}. The original VerilogEval infrastructure features a monolithic \texttt{JSONL} file with the dataset and a Python script to evaluate, while the proposed v2 flow uses a Makefile-based approach. A Makefile parameter specifies which of multiple datasets is used for the evaluation, and each dataset contains the problem prompts, reference solutions, and in-context learning examples. Two datasets are included in the VerilogEval v2 repository, one for code completion and one for specification-to-RTL tasks. An evaluation Python script, similar to the one in  \cite{liu2023verilogeval}, queries the LLM under test with prompts contains problem descriptions along with in-context learning examples, if applicable. LLM responses are saved in a working directory specific to each problem. Icarus Verilog is then used to evaluate the generated SystemVerilog against the reference solution. Lastly, a failure classification script detects keywords in the Icarus Verilog output (both compile time and runtime) to classify failures. A summary of failures across problems is saved into a text file for human analysis.

\subsection{Specification-to-RTL Task Support}

The enhanced VerilogEval v2 benchmark supports both code completion and specification-to-RTL tasks to better match the instruction-tuning \cite{yuan2023evaluating} of recent models.  The full 156-problem dataset from VerilogEval is converted into specification-to-RTL prompting in this work. Code completion has the problem description in Verilog-compatible comments and always appends the module interface declaration to the end of the prompt, similar to how Copilot \cite{copilot_github_2024} is typically implemented in an integrated development environment (IDE).

On the other hand, specification-to-RTL's prompt style is like a chatbot, with well-defined ``Question'' and ``Answer'' sections. The specification-to-RTL prompting is implemented in a manner similar to the Mostly Basic Python Problems (MBPP) benchmark\cite{austin2021program} with \texttt{[BEGIN]} and \texttt{[DONE]} tags surrounding code blocks. Examples of these two styles can be found in listings~\ref{lst1:1shot_example_code_completion} and \ref{lst1:1shot_example_spec_to_rtl} with only the highlighted code indicating the prompt styles.

\subsection{Support for In-Context Learning Examples}

In-context learning (ICL) was proposed by \cite{brown2020language} to add examples of task questions and desired responses into the prompt context, so that an LLM can better respond to a given task. ICL is implemented through simple Verilog code examples, tailored for both code completion (Listing~\ref{lst1:1shot_example_code_completion}) and specification-to-RTL tasks (Listing~\ref{lst1:1shot_example_spec_to_rtl}). The listings contain the 1-shot examples used for both tasks, except line width and whitespace were adjusted for printing. The examples were selected to be short and simple, while including a full module (from declaration to \texttt{endmodule}).

Two additional examples for each task are also added to the infrastructure, as shown in listings \ref{lst1:2shot} and \ref{lst1:3shot}: a sequential incrementer (Listing~\ref{lst1:2shot}) similar to the first 1-shot example, and a basic finite-state machine (Listing~\ref{lst1:3shot}). The number of shots is parameterized and can easily be swept to determine sensitivity of a model's pass rate as ICL examples are added to the prompt. 1-shot includes only the combinational incrementer, 2-shot adds the sequential incrementer, and 3-shot includes all three examples in the context prompt.


\lstset{frame=topline}
\begin{lstlisting}[
  caption=The 1-shot in-context learning example for code completion tasks. The highlighted code is the prompt style.,
  label=lst1:1shot_example_code_completion,
  language=verilog,
  belowskip=0pt,backgroundcolor=\color{yellow}]
// Implement the Verilog module based on the
// following description. Assume that sigals
// are positive clock/clk triggered unless
// otherwise stated.
//
// The module should implement an incrementer
// which increments the input by one and
// writes the result to the output. Assume
// all values are encoded as two's complement
// binary numbers.
module TopModule
(
  input  logic [7:0] in_,
  output logic [7:0] out
);
\end{lstlisting}
\lstset{frame=bottomline}
\begin{lstlisting}[language=verilog,aboveskip=0pt]
  // Combinational logic
  assign out = in_ + 1;
endmodule
\end{lstlisting}

\lstset{frame=topline}
\begin{lstlisting}[
  caption=The 1-shot ICL example for specification-to-RTL tasks. The highlighted code is the prompt style.,
  label=lst1:1shot_example_spec_to_rtl,
  language=,
  belowskip=0pt,backgroundcolor=\color{yellow}
]
Question:
Implement a hardware module named TopModule
with the following interface. All input and
output ports are one bit unless otherwise
specified.

 - input  in_ (8 bits)
 - output out (8 bits)

The module should implement an incrementer
which increments the input by one and writes
the result to the output. Assume all values
are encoded as two's complement binary
numbers.

Enclose your code with [BEGIN] and [DONE].
Only output the code snippet and do NOT output
anything else.

Answer:
\end{lstlisting}
\lstset{frame=bottomline}
\begin{lstlisting}[language=,aboveskip=0pt]
[BEGIN]
module TopModule
(
  input  logic [7:0] in_,
  output logic [7:0] out
);
  // Combinational logic
  assign out = in_ + 1;
endmodule
[DONE]
\end{lstlisting}

\lstset{frame=topline|bottomline}
\begin{lstlisting}[
  caption=The 2-shot in-context learning example for code completion tasks.,
  label=lst1:2shot,
  language=verilog]
// Implement the Verilog module based on
// the following description. Assume that sigals
// are positive clock/clk triggered unless
// otherwise stated.
//
// The module should implement an 8-bit registered incrementer with an
// active-high synchronous reset. The 8-bit input is first registered and
// then incremented by one on the next cycle. The internal state should be
// reset to zero when the reset input is one. Assume all values are encoded
// as two's complement binary numbers. Assume all sequential logic is
// triggered on the positive edge of the clock.

module TopModule
(
  input  logic       clk,
  input  logic       reset,
  input  logic [7:0] in_,
  output logic [7:0] out
);
  // Sequential logic
  logic [7:0] reg_out;
  always @( posedge clk ) begin
    if ( reset )
      reg_out <= 0;
    else
      reg_out <= in_;
  end
  // Combinational logic
  logic [7:0] temp_wire;
  always @(*) begin
    temp_wire = reg_out + 1;
  end
  // Structural connections
  assign out = temp_wire;
endmodule
\end{lstlisting}

\lstset{frame=topline|bottomline}
\begin{lstlisting}[
  caption=The 3-shot in-context learning example for code completion tasks.,
  label=lst1:3shot,
  language=verilog]
// Implement the Verilog module based on
// the following description. Assume that sigals
// are positive clock/clk triggered unless
// otherwise stated.
//
// Build a finite-state machine that takes as input a
// serial bit stream and outputs a one whenever the bit stream contains two
// consecutive one's. The output is one on the cycle _after_ there are two
// consecutive one's.
//
// The reset input is active high synchronous and should reset the
// finite-state machine to an appropriate initial state.
module TopModule
(
  input  logic clk,
  input  logic reset,
  input  logic in_,
  output logic out
);
  // State enum
  localparam STATE_A = 2'b00;
  localparam STATE_B = 2'b01;
  localparam STATE_C = 2'b10;
  // State register
  logic [1:0] state;
  logic [1:0] state_next;
  always @(posedge clk) begin
    if ( reset )
      state <= STATE_A;
    else
      state <= state_next;
  end
  // Next state combinational logic
  always @(*) begin
    state_next = state;
    case ( state )
      STATE_A: state_next = ( in_ ) ? STATE_B : STATE_A;
      STATE_B: state_next = ( in_ ) ? STATE_C : STATE_A;
      STATE_C: state_next = ( in_ ) ? STATE_C : STATE_A;
    endcase
  end
  // Output combinational logic
  always @(*) begin
    out = 1'b0;
    case ( state )
      STATE_A: out = 1'b0;
      STATE_B: out = 1'b0;
      STATE_C: out = 1'b1;
    endcase
  end
endmodule
\end{lstlisting}

\subsection{Support for Failure Classification}

Failures of LLM-generated responses are automatically classified by broad reasons for failure, both Verilog compile time errors and simulation runtime errors, such as incorrectly using a wire as a register, incorrect bit widths, and missing module interface definitions. This classification feature provides insight into the most common reasons for failures and how to mitigate poor code generation through prompt tuning. The classification is dependent on specific warnings and errors given by Icarus Verilog or the test harness. The failures are classified in Table~\ref{tab:failure_types}.

Classifications were developed by human inspection of common failure modes across the code completion benchmark. For example, LLMs were observed frequently mixing up the use of registers and wires. Solutions in prompt tuning could vary: from adding prompt rules to only use wires on ports to suggesting the use of SystemVerilog \texttt{logic} port types, obviating the immediate type confusion, to allowing the LLM to generate the interface entirely on its own (as in the case of specification-to-RTL, rather than code completion). By classifying failures, the impact of prompt changes on code generation performance can be directly observed and guided.

\begin{table}[htbp]
    \centering
    \caption{Types of failures supported by automatic failure classification.}
    \begin{tabular}{|p{0.25\linewidth}|p{0.65\linewidth}|}
        \hline
        \textbf{Failure Type} & \textbf{Example} \\
        \hline
        \hline
        \multicolumn{2}{|l|}{\textbf{Compile-Time Failures}} \\
        \hline
        \raggedright
        Unable to Bind Wire/Reg 'clk' & Clk is missing in interface ports list, such as if a code completion task does not specify a clock to be used yet the LLM used it in the generated code. \\
        \hline
        \raggedright
        Unable to Bind Wire/Reg & Other port related bind problems.\\
        \hline
        \raggedright
        Explicit Cast Required & A datatype problem occurred, often with use of enums.\\
        \hline
        \raggedright
        Module Missing & Typically indicates the modular declaration is missing from the generated code. \\
        \hline
        \raggedright
        Sensitivity Problem & Sensitivity lists for always blocks are not defined properly. \\
        \hline
        \raggedright
        Reg Declared as Wire & A wire is assigned to as a reg. \\
        \hline
        \raggedright
        Syntax Error & General syntax errors in generated code. \\
        \hline
        \raggedright
        General Compiler Error & Other compiler errors without specific classification. \\
        \hline
        \hline
        \multicolumn{2}{|l|}{\textbf{Run-Time Failures}} \\
        \hline
        \raggedright
        Reset Issue & Reset should be synchronous but is asynchronous. \\
        \hline
        \raggedright
        Timeout & The simulation did not complete in reasonable time, indicating a sequential block does not have a correct implementation. \\
        \hline
        \raggedright
        General Runtime Error & Other runtime errors that are not classified, including mismatched outputs. \\
        \hline
    \end{tabular}
    \label{tab:failure_types}
\end{table}

\subsection{Other Infrastructural Improvements}

The original VerilogEval benchmark contained all problems in a monolithic \texttt{JSONL} format. This is efficient to run, but inefficient to inspect manually using a text editor. In the improved benchmark, each problem was split into a set of files, including problem prompts, module interfaces, and test benches. Autoconf\cite{noauthor_autoconf_nodate} and GNU Make\cite{noauthor_make_nodate} were employed to target a model evaluation build directory to a specific evaluation target, including the LLM itself to run, number of shots, number of samples, task to complete, and other parameters. For each problem, a resulting problem evaluation directory is created containing a log of the LLM prompt/responses, generated Verilog file, and the Icarus Verilog output log. This infrastructure allows for scalable sweeps through the use of Make's parallel run feature, helps continue an evaluation run if it is interrupted, and allows for easy human inspection of the resulting collateral. A backwards-compatible mode with \texttt{JSONL} support is planned for VerilogEval v2.

\vspace{3mm}

\section{VerilogEval v2 Evaluation}



\begin{figure*}[hbtp]
    \centering
    \subfigure[Code Completion Task]{
    \includegraphics[scale=0.48,trim={10cm 3cm 12.5cm 3cm},clip]{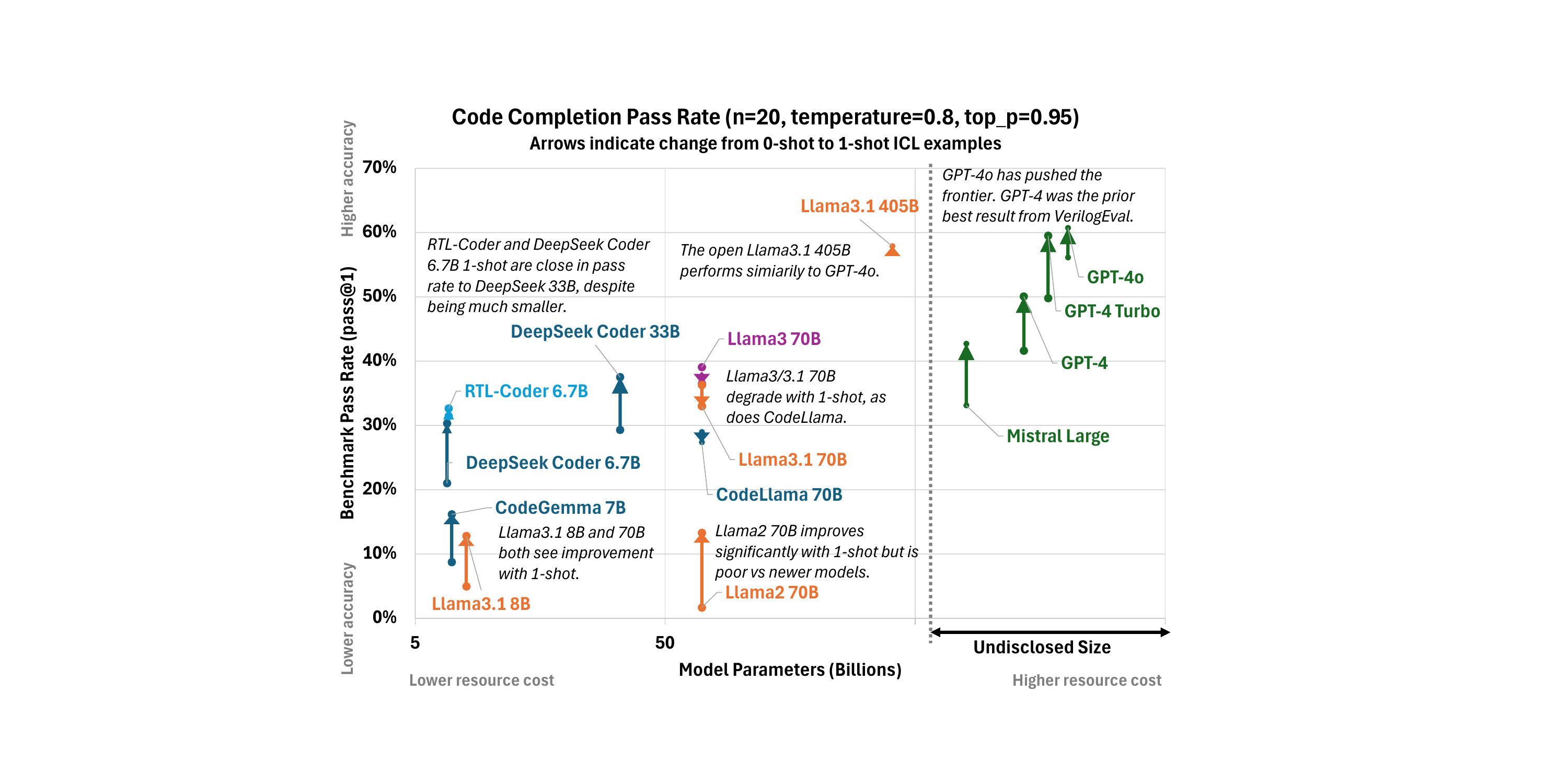}
    \label{fig:pass_rates:code_completion}}\quad
    \subfigure[Specification-to-RTL Task]{
    \includegraphics[scale=0.48,trim={10cm 3cm 12.5cm 3cm},clip]{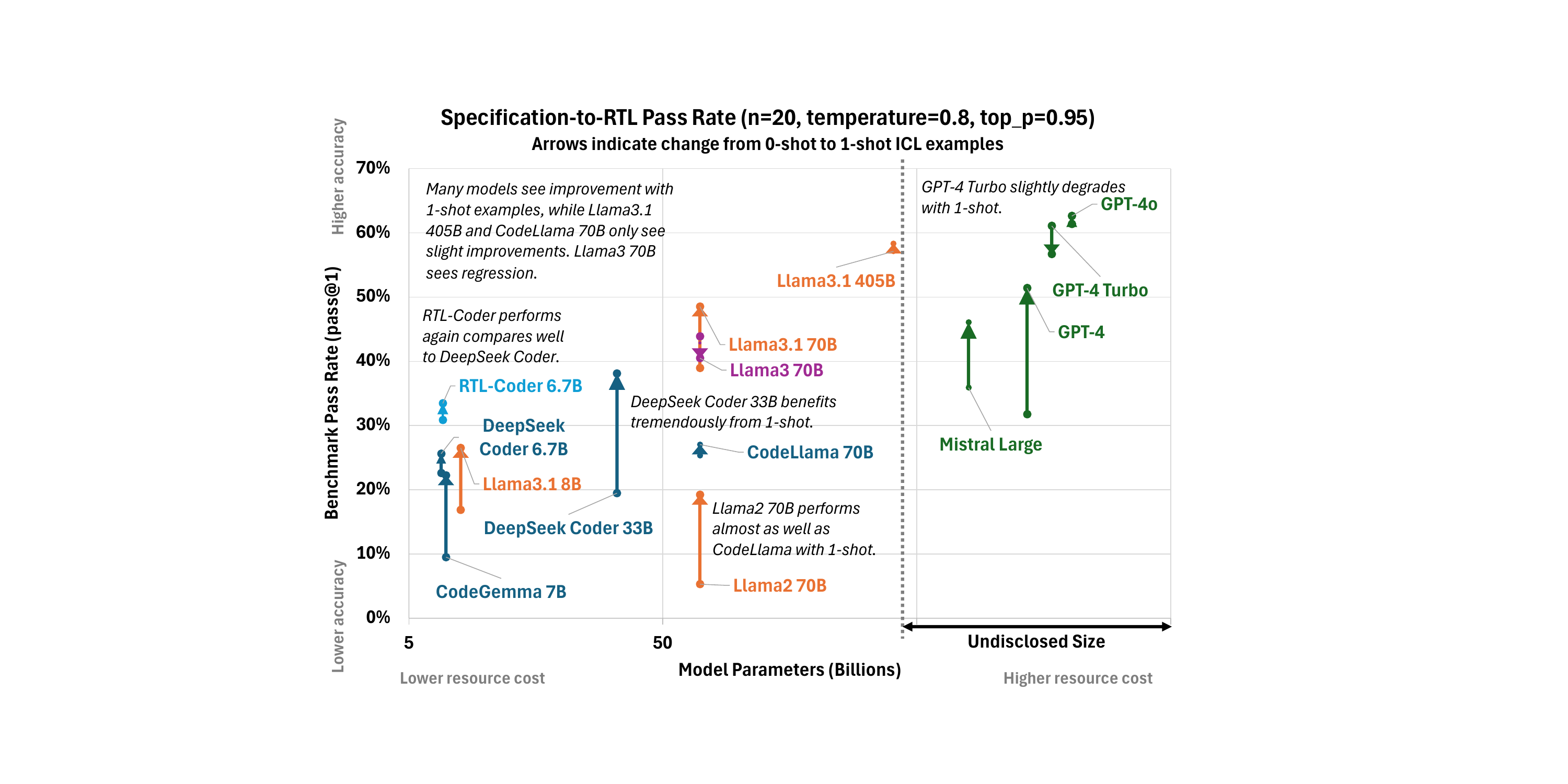}
    \label{fig:pass_rates:spec-to-rtl}}
    \caption{Pass rate across recent large-language models. Green models are closed general-purpose models, orange are open general-purpose models, dark blue are coding-specific models, and light blue is an RTL-specific model. Purple is the older Llama3~70B to demonstrate a large degradation due to ICLs.}
    \label{fig:pass_rates}
\end{figure*}

The graph in Figure~\ref{fig:pass_rates} illustrates the performance of the recent large-language models (LLMs) on code completion and specification-to-RTL translation tasks, as measured by the benchmark pass rate (pass@1 in \cite{liu2023verilogeval}). As in the previous section, model results were captured as both a 20-sample (n=20) high-temperature (T=0.8, top\_p=0.95) set and 1-sample (n=1) low-temperature (T=0.0, top\_p=0.01) set.  Models are arranged along the x-axis by model size, with undisclosed model sizes on the right. The evaluation compares models with and without 1-shot in-context learning (ICL) examples, represented by arrows indicating the change in performance as 1-shot examples are added. For code completion tasks, Llama3.1~405B initially achieves the highest pass rate in 0-shot, as previously shown in Figure~\ref{fig:survey}. However, when 1-shot is added, GPT-4o achieves the highest pass rate at approximately 61\% from 56\% in 0-shot, establishing the new state-of-the-art frontier. As both prompt configurations show, GPT-4o has robust improvement over GPT-4 for RTL generation tasks.


Llama3.1~405b established the new Pareto frontier at 0-shot (57.0\%), demonstrating that open models have matched closed commercial models, but showed very little improvement in 1-shot (57.9\%). The older Llama3~70B (purple line) was included to demonstrate a clear counter example to ICL's improving pass rate. While Llama3.1 generally improves with in-context learning examples, Llama3~70B declines in pass rate when the 1-shot ICL example is added to the prompt, which will be discussed in detail in the next section. Among the smaller specialized models, RTL-Coder~6.7B showed an impressive pass rate of around 33\%, while being much smaller than general-purpose models. RTL-Coder when originally sampled did not properly insert whitespace after \texttt{endmodule} statements and would often repeat code blocks. We modified our post-process script that extracts the Verilog code from the response to match the post-processing in RTL-Coder's evaluation scripts \cite{software-hkust-zhiyao_hkust-zhiyaortl-coder_2024}, and Figure~\ref{fig:pass_rates}'s RTL-Coder results are shown using this modified response extraction. This post-processing is also used across the other models as well. DeepSeek Coder 6.7B (30.3\%) nearly reaches the pass rate of RTL-Coder (32.6\%) in code completion when an ICL example is added.

Specification-to-RTL task results showed generally similar pass rates compared to code completion, with some exceptions. GPT-4 Turbo showed noticeable pass rate improvement in code completion tasks, but some degradation in spec-to-RTL. Mistral Large showed improvements in both tasks. Llama3.1~70B and CodeGemma~7B saw much larger improvements in specification-to-RTL when adding 1-shot ICL. In particular, at this prompt tuning configuration, Llama3.1 70B exceeds the highest pass rate achieved by Llama3 70B across all configurations presented, despite Llama3.1 starting at a lower pass rate than Llama3 across both tasks at 0-shot. In Llama3.1~405B across both tasks, adding an ICL example made little difference in pass rate. Llama3~70B saw decline with 1-shot in spec-to-RTL, more so than in code completion. As with code completion, RTL-Coder 6.7B (33.5\%) maintained a lead over DeepSeek Coder 6.7B in spec-to-RTL (25.6\%). 

The full results are shown in Table~\ref{tab:pass_rates} and include both n=20 (20 samples, temperature=0.8, top\_p=0.95) from Figure~\ref{fig:pass_rates} along with deterministic n=1 (1 sample, temperature=0.0, top\_p=0.01). Some models performed notable better in spec-to-RTL than code completion, such as Llama3.1 70B 1-shot with 48.5\% in code completion versus 39.0\% in spec-to-RTL. This variability underscores the importance of tailored prompt tuning and the potential of ICL to enhance code generation performance in certain models.

\begin{table*}[htbp]
\caption{VerilogEval Pass Rates of Recent Large-Language Models. Number of samples $n=1$ when $T=0$ and $n=20$ when $T=0.8$.}
\begin{center}
\small
  \begin{tabular}{|c|c|c|c|*{2}{|cc|cc|}}
    \hline
    \textbf{Model Name} & \textbf{Model Size} & \textbf{License} & \textbf{Type} & \multicolumn{4}{|c||}{\textbf{Task: Code Completion}} & \multicolumn{4}{|c|}{\textbf{Task: Specification-to-RTL}} \\
    \hline
    \hline
     \multicolumn{4}{|r||}{\textit{In-Context Learning Examples:}} & \multicolumn{2}{|c|}{\textbf{0-Shot}} & \multicolumn{2}{|c||}{\textbf{1-Shot}} & \multicolumn{2}{|c|}{\textbf{0-Shot}} & \multicolumn{2}{|c|}{\textbf{1-Shot}} \\
     \multicolumn{4}{|r||}{\textit{Temperature:}} & \textbf{T=0} & \textbf{T=0.8} & \textbf{T=0} & \textbf{T=0.8} & \textbf{T=0} & \textbf{T=0.8} & \textbf{T=0} & \textbf{T=0.8}\\
    \hline

    GPT-4o\cite{gpt4o} & Undisclosed & Closed & General & 59.0\% & 56.1\% & 62.8\% & 60.7\% & 62.5\% & 61.4\% & 65.1\% & 62.6\% \\
    GPT-4 Turbo\cite{gpt4_turbo_announce} & Undisclosed & Closed & General & 53.9\% & 49.8\% & 59.6\% & 59.5\% & 59.6\% & 61.1\% & 56.4\% & 56.7\% \\
    GPT-4\cite{gpt4} & Undisclosed & Closed & General & 42.3\% & 41.6\% & 51.3\% & 50.1\% & 32.0\% & 31.7\% & 48.7\% & 51.4\% \\
    Mistral Large\cite{ai_au_2024} & Undisclosed & Closed & General & 34.0\% & 33.1\% & 44.2\% & 42.7\% & 37.5\% & 35.9\% & 48.7\% & 46.0\% \\
    Llama3.1\cite{noauthor_meta-llamallama3_2024} & 405B & Open & General & 56.4\% & 57.0\% & 59.6\% & 57.9\% & 57.2\% & 57.1\% & 57.9\% & 58.3\% \\
    \hline
    Llama3.1\cite{noauthor_meta-llamallama3_2024} & 70B & Open & General & 35.3\% & 36.3\% & 34.0\% & 33.0\% & 42.8\% & 39.0\% & 48.0\% & 48.5\% \\
    Llama3\cite{noauthor_meta-llamallama3_2024} & 70B & Open & General & 37.8\% & 39.1\% & 36.5\% & 36.5\% & 40.8\% & 43.9\% & 39.5\% & 40.5\% \\
    Llama2\cite{noauthor_meta-llamallama3_2024} & 70B & Open & General & 1.3\% & 1.7\% & 15.4\% & 13.3\% & 4.6\% & 5.3\% & 17.8\% & 19.2\% \\
    CodeLlama\cite{noauthor_meta-llamacodellama-70b-instruct-hf_2024} & 70B & Open & Coding & 37.2\% & 29.0\% & 41.7\% & 27.4\% & 34.9\% & 25.3\% & 41.5\% & 27.0\% \\
    DeepSeek Coder\cite{guo2024deepseekcoder} & 33B & Open & Coding & 25.0\% & 29.3\% & 42.3\% & 37.5\% & 21.7\% & 19.5\% & 40.1\% & 38.1\% \\
    \hline
    Llama3.1\cite{noauthor_meta-llamallama3_2024} & 8B & Open & General & 2.6\% & 4.9\% & 10.9\% & 12.8\% & 19.1\% & 16.8\% & 27.6\% & 26.5\% \\
    CodeGemma\cite{noauthor_googlecodegemma-7b_2024} & 7B & Open & Coding & 8.3\% & 8.7\% & 19.9\% & 16.2\% & 6.6\% & 9.5\% & 24.3\% & 22.2\% \\
    DeepSeek Coder\cite{guo2024deepseekcoder} & 6.7B & Open & Coding & 24.4\% & 21.0\% & 33.3\% & 30.3\% & 29.6\% & 22.6\% & 27.6\% & 25.6\% \\
    RTL-Coder\cite{liu2024rtlcoder} & 6.7B & Open & Verilog RTL & 35.9\% & 31.5\% & 37.2\% & 32.6\% & 36.8\% & 30.9\% & 34.9\% & 33.5\% \\
    \hline
  \end{tabular}
\label{tab:pass_rates}
\end{center}
\end{table*}

Overall, larger models generally achieve higher pass rates, though resource costs and model-specific responses to ICL examples vary significantly. Within the context of VerilogEval, GPT-4o and Llama3.1~405B have become clear leaders for the highest achieved pass rates, demonstrating that open models (Llama3.1~405B) have reached parity with closed models. Additionally, smaller (70B) open models have become competitive with last year's larger closed models. Domain-specific models (RTL-Coder) are also competitive in some scenarios at a much smaller size.
\vspace{3mm}

\section{Impact of ICL on Pass Rates and Failures}

As demonstrated in the previous section, in-context learning examples improve model generation accuracy in some conditions but degrade accuracy in others. ICL impact bears further investigation to better understand strategies to apply prompt tuning.

\subsection{Increased In-Context Learning Examples}
\begin{figure}[htbp]
    \centering
    \includegraphics[trim=10cm 3cm 10cm 3cm,clip,scale=0.48]{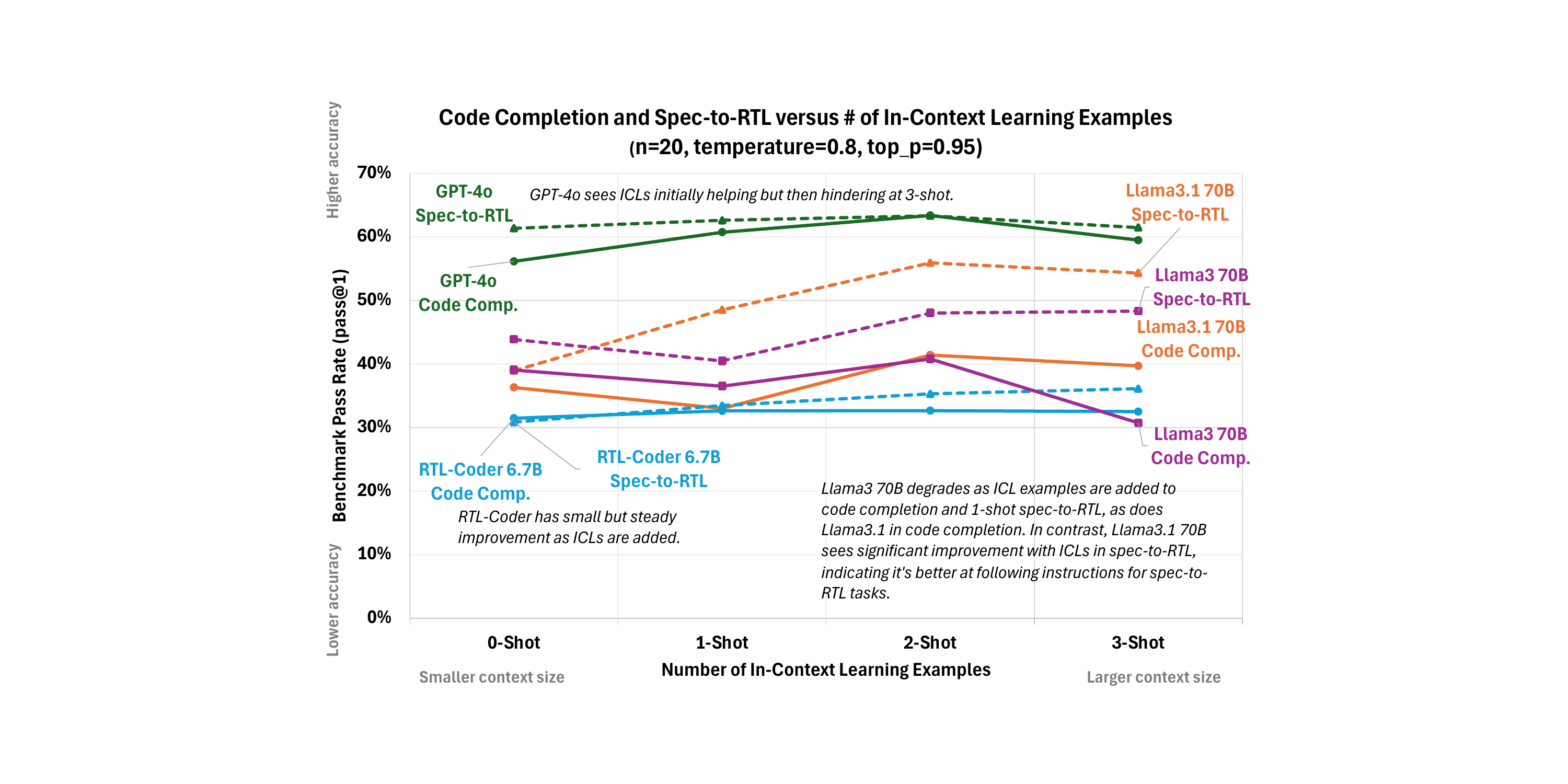}
    \caption{Pass rate of three models for code completion and specification-to-RTL tasks, with 0-shot to 3-shot in-context learning examples. Solid lines are code completion and dashed lines are spec-to-RTL.}
    \label{fig:pass_rate_vs_icl}
\end{figure}

Higher-shot ICL runs were conducted for four models across parameter size classes: GPT-4o, Llama3.1 70B, Llama3 70B, and RTL-Coder 6.7B. Pass rates of these four models for the two tasks across 0-shot to 3-shots are shown in Figure~\ref{fig:pass_rate_vs_icl}. The figure highlights the varying impact of ICL examples on different models and tasks, emphasizing the potential benefits of task-specific tuning and the necessity of providing contextual examples to enhance model outputs. Notably, GPT-4o exhibits stable and high performance across all ICL example counts of at least 1-shot, maintaining a pass rate of 55\% to 63\%. In contrast, Llama3 70B demonstrates divergent trends: its spec-to-RTL performance improves from 40\% to nearly 50\% with more ICL examples, whereas its code completion performance declines from 40\% to just above 30\%. Llama3.1 70B achieves even better pass rates with ICL examples as compared to Llama3 for spec-to-RTL, and is fairly stable in code completion. RTL-Coder shows stability from 0-shot to 3-shot in both code completion and spec-to-RTL, only in very little improvements in the latter case. To understand how specific in-context learning examples can influence response pass and failure, we will look at specific case examples in the next section across Llama models. 

\subsection{Case Study: Problem 9 and Problem 34}

In-context learning may improve model pass rate through better aligning the model response with the task at hand if the ICL example is well-chosen. To better understand the impact of in-context learning on Verilog code generation pass rate, we consider two problems from the dataset: Problem 9 and Problem 34. Problem 9 is a simple combinational design question that asks: ``\texttt{A "population count" circuit counts the number of '1's in an input vector. Build a population count circuit for a 3-bit input vector.}'' In contrast, Problem 34 is a simple sequential design question: ``\texttt{Create 8 D flip-flops. All DFFs should be triggered by the positive edge of clk.}''

In 0-shot (no in-context learning examples are given) for Problem 9, zero of the 20 high-temperature samples pass on Llama3 70B. The failure case is an ``\texttt{Unable to Bind Wire/Reg `clk'}'' error given by Icarus Verilog, in which the LLM tries to use a nonexistent clock signal. In effect, the LLM is confusing a combinational design problem for a sequential design problem. An example failing response to Problem 9 is given in the listing below:

\lstset{frame=single}
\begin{lstlisting}[language=verilog]
module TopModule (
  input [2:0] in,
  output [1:0] out
);
  reg [1:0] out_reg;
  always @(posedge clk) begin
      out_reg = (in[2] == 1'b1) + (in[1] == 1'b1) + (in[0] == 1'b1);
  end
  assign out = out_reg;
endmodule
\end{lstlisting}

In this example, the clock is not present, which causes the error. Additionally, as can be seen from the example, the output port is assumed to be registered when instead it is an implicit wire. In code completion tasks, the LLM does not have the flexibility to define the port types (\texttt{wire} or \texttt{reg}), but attempts to use the port as an incorrect type. When the 1-shot example shown in Listing~\ref{lst1:1shot_example_code_completion}, which demonstrates a simple combination incrementer, is added to the prompt as an in-context learning example, the pass rate of Problem 9 increases substantially to 65\% (13 of the 20 samples pass). A typical Llama3 70B passing response to this problem is:

\begin{lstlisting}[language=verilog]
module TopModule (
  input [2:0] in,
  output [1:0] out
);
  // Combinational logic
  assign out = (in[2] + (in[1] + in[0]));
endmodule
\end{lstlisting}

In effect, the LLM response has been better aligned to handle combinational design problems, at least for a trivial example like Problem 9. Of the remaining 7 failures in 1-shot, three are missing an \texttt{endmodule} keyword that terminates the module definition, and four try to use the output port as a register instead of a wire (\texttt{Reg Declared as Wire}). This suggests that even with the ICL example, the LLM does not understand where it should terminate its response and the distinction between wires and registers.

The added 1-shot example focuses on combinational design, yet Problem 34 is a sequential design. Initially, the 20-sample pass rate of Problem 34 is 100\% in 0-shot but drops significantly to 5\% in 1-shot. Thus, a question arises of whether sequential design problems are improved in 1-shot at the expense of combinational design problems. However, on inspection, all failing responses are due to a missing \texttt{endmodule} and the above-mentioned confusion on sequential versus combinational design. An example failing response for Problem 34 is, where an \texttt{endmodule} keyword is missing:

\begin{lstlisting}[language=verilog]
module TopModule (
  input clk,
  input [7:0] d,
  output reg [7:0] q
);

always @(posedge clk)
  q <= d;
\end{lstlisting}

 Adding another ICL example in 2-shot (Listings~\ref{lst1:2shot}, a combinational design example) increases the Problem 9 pass rate to 100\% in 2-shot and 3-shot, eliminating both of the above-mentioned missing \texttt{endmodule} and reg versus wire confusion demonstrated for Problem 9. However, only 50\% of the 20 sample responses pass for Problem 34 in 2-shot, with all failing cases missing an \texttt{endmodule}. One culprit may be the confusion with \texttt{begin} and \texttt{end} block start and end keywords. All failing cases use \texttt{being} and \texttt{end} for the sequential \texttt{always} block, while in the passing cases \texttt{begin} and \texttt{end} are omitted. 

 Adding one more ICL example in 3-shot (Listing~\ref{lst1:3shot}, an FSM design example), Problem 34 gets even worse: 0\% of the cases pass. All cases use \texttt{begin} and \texttt{end} while omitting \texttt{endmodule}. Ostensibly, the multiple always of Listing~\ref{lst1:3shot} caused the LLM to over-emphasize the \texttt{always} blocks with \texttt{begin} and \texttt{end}, but did not learn how to correctly use \texttt{endmodule}. However, Problem 9 maintains a 100\% pass rate in 3-shot.

 While the code completion task by nature constrains the port types (\texttt{reg} or \texttt{wire}), specification-to-RTL allows the full interface specification to be given in the response. This leads to a higher 0-shot pass rate on Llama3 with 75\% on Problem 9 (compared to 0\% in code completion) and 100\% pass rate on Problem 34 (compared to 65\%).  However, in 1-shot the spec-to-RTL pass rate drops to 0\% on Problem 9 because of it picking up a slightly different input port name (\texttt{in\_}) used in the in-context learning example that is not present in the prompt. Problem 34 is unaffected and is successful across all samples. Increasing ICL examples to 2-shot fully recovers and surpasses 0-shot to 100\% pass rates across both problems. Similarly, 3-shot has full 100\% pass rates on both. Overall, specification-to-RTL tasks perform better than code completion for these problems on Llama3 70B.

Broadening the results across older and newer generation of the Llama 70B models on code completion, Llama2 70B only has a 0\% and 15\% pass rate for Problem 9 and 34, respectively, with an assortment of failures. In 1-shot, Problem 9 shows no improvement, while Problem 34 increases to 30\%. Code sections are often repeated in the responses, causing failures and demonstrating poor instruction following. Adding additional ICL examples continues to show no improvement, with Problem 9 never passing in 2- or 3-shot and Problem 34 regressing to 15\% in 2-shot and 5\% in 3-shot.

The newer Llama3.1 70B in 0-shot fails all Problem 9 samples, from a mix of failures including using a non-existent clock and redefining the output port as a register, but passes all Problem 34. Adding 1-shot, Problem 9 improves to 75\% pass and Problem 34 maintain 100\% pass across all 20 samples. Interestingly, all five failures in Problem 9 are due to bad combinational logic for the counting task, causing mismatch against the golden reference in simulation, and not compile-time errors. In particular, the boolean expressions derived for population counting are incorrect.

Adding the Listing~\ref{lst1:2shot} ICL example to the prompt in 2-shot to Llama3.1 70B code completion reduces Problem 9 pass rates to 60\% and causes all sample responses on Problem 34 to fail. In this case, the failure is consistent inclusion of English natural language explanation of the model being created, causing obvious syntax issues. None of the Problem 9 failures feature this failure mode. Increasing ICL examples to 3-shot completely clears up the Problem 34 failures and returns Problem 9 to 75\% with only boolean logic failures. As we can see for Llama3.1 70B, adding in-context learning examples can generally improve results, but sometimes wrong behavior will be learned, as in the case of 2-shot.

The much larger Llama3.1 405B, exhibited 100\% 20-sample pass rates for Problem 9 and Problem 34 in all ICL shot scenarios, understanding these basic problems well. Thus, the exact impact of prompt tuning on a model can be very model dependent in terms of model parameter size, lineage, and generation.

\subsection{Aggregate Failure Analysis}

\begin{figure*}
    \centering
    \includegraphics[trim=5.5cm 1cm 5.5cm 1cm,clip,scale=0.38]{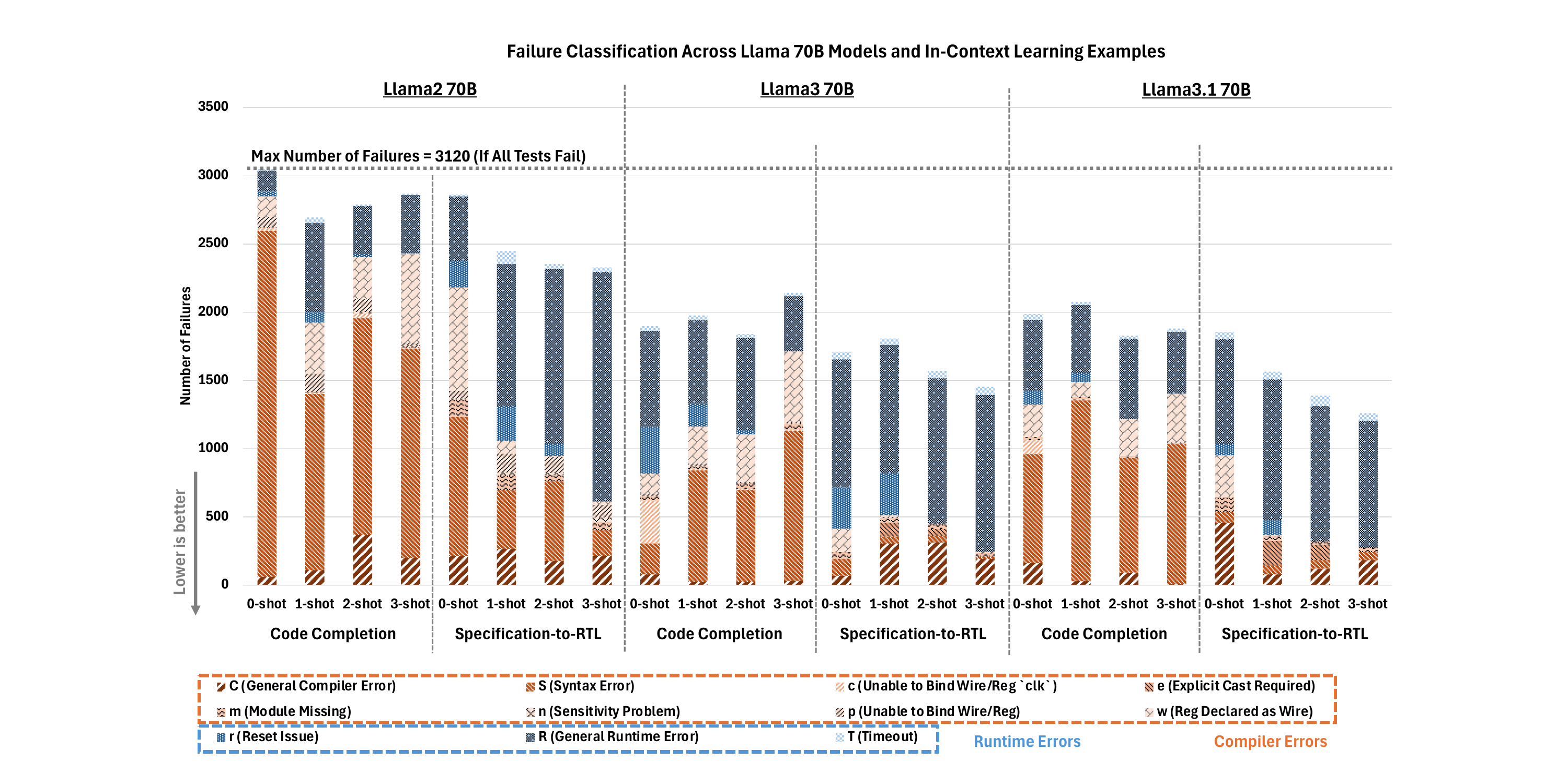}
    \caption{Failure classification for Llama2 70B, Llama3 70B, and Llama3.1 70B models with 0-shot to 3-shot ICL examples across the two tasks. Orange coloring indicates compiler errors, while blue indicates runtime issues.}
    \label{fig:icl_failures}
\end{figure*}

Figure~\ref{fig:icl_failures} employs the new failure classification feature of the improved benchmark infrastructure to illustrate the number and types of failures encountered by different models across various numbers of in-context learning (ICL) examples. The y-axis represents the number of failures, with lower values indicating better pass rates. Each bar is segmented to show different categories of errors, with orange shades representing compiler errors and blue shades representing runtime errors. The figure is divided into three sections for three generations of Llama 70B models, highlighting the numbers and types of failures across 0-shot to 3-shot ICL examples. As compiler errors will be flagged and mask runtime errors (since code that does not compile never runs), the bars on the graph are best read from bottom to top. A reduction in runtime errors for the same total bar height indicates that compiler errors have displaced runtime errors. This layering effect should be considered when interpreting the improvements or degradations in model performance as additional ICL examples are introduced.

Llama2 70B initially improves in code completion from 0-shot to 1-shot, but then degrades as more examples are added. The ``\texttt{Reg Declared as Wire}'' error which confuses wire output ports with registered output ports, as discussed in the last section, is especially pronounced at 3-shot. In comparison, specification-to-RTL greatly improves compile-time errors, including wire and registered output confusion, even at 1-shot. This suggests that specification-to-RTL has a naturally higher pass rate because of the flexibility of the LLM to define the interface types. However, runtime errors remain as the LLM is unable to solve the dataset problems with correct logic. 

Llama3 and Llama3.1 70B both exhibit a similar pattern of code completion versus specification-to-RTL, with ``\texttt{Reg Declared as Wire}'' errors mostly eliminated in high ICL shot cases. However, only Llama3.1 70B improves substantially and consistently as ICL examples are added. It's also notable that general syntax errors are much reduced in specification-to-RTL as compared to code completion. This is true even in 0-shot cases, although the overall failure rate is similar. In the context of code completion in Llama3 70B, the inclusion of ICL examples has a tendency to increase the number of compile-time failures, whereas the failure rate remains relatively stable with ICL examples in Llama3.1 70B.

The results emphasize the need for careful tuning of ICL examples to optimize results.  While ICL can help correct certain types of mistakes, it can also introduce new issues, leading to similar or even worse performance. In addition to the failure classification feature capturing high-level counts of types of failures across different models and prompting settings, it also allows for detailed inspection on a problem-by-problem basis within a run. This granular analysis helps identify whether specific problems or categories of problems have systematic types of failures. Such insights can guide more careful tuning of prompts across the benchmark, leading to more effective and targeted improvements in model performance. A careful analysis of the problem categories within VerilogEval and comparative failure counts could help find the best ICL examples to use for a given model.








\vspace{3mm}

\section{Future Work}

Since its release last year, VerilogEval has been commonly used in state-of-the-art LLM Verilog code generation research, with over 100 citations\footnote{As reported by Google Scholar.}. However, application of VerilogEval is quickly becoming limited due to the low complexity of the dataset problems, especially when LLM deployments move beyond single-turn prompts and responses and into multi-turn flows with feedback, as we shall see below. Additionally, hardware design extends far beyond specification-to-RTL code generation tasks and benchmarks need to cover many other tasks for LLM-based hardware design automation to continue.

\subsection{Agent-based Code Generation}

Moving beyond single-turn prompt-response into agent-based approaches is key to solving more complex tasks. LLM-based \textit{agentic} approaches are typically composed of a high-level planning agent (a dedicated prompt and model pair that is optimized for high-level planning) to enumerate tasks to prompted problem, and dedicated expert LLMs to implement said tasks. Additionally, there may be LLMs to integrate and combine disparate responses, ranking the best responses, and detecting if a given problem is solved or if instead the solution must be refined further. These dedicated LLMs may use the same LLM model but with different prompting or context, or models may be heterogeneous to efficiently solve targeted tasks.


Recent agent-based Verilog code generation approaches include RTLFixer \cite{tsai-rtlfixer-arxiv2023}, VeriAssist \cite{huang2024llmpoweredverilogrtlassistant},  AIvril \cite{islam2024aivrilaidrivenrtlgeneration}, MAGE \cite{zhao2024magemultiagentengineautomated}, PromptV \cite{mi2024promptvleveragingllmpoweredmultiagent}, and VerilogCoder \cite{ho2024verilogcoder}. RTLFixer employs RAG and a thought-action-observation loop to resolve syntactical issues found during code compilation, improving VerilogEval-human $pass@1$ by 10\%. VeriAssist applies multi-turn, chain-of-thought reasoning to fix both syntactical and functional errors in generated code, and improves VerilogEval-human $pass@1$ by about 7\%. AIvril includes a coding agent to generate code, and a review agent to analyze compilation and simulation errors. The review agent provides feedback to the coding agent to revise the code. Their approach improves $pass@1$ by a 14\% increase to a total 65\% on VerilogEval when using gpt-4o. PromptV employs code and testbench generation agents, code and testbench learner agents, and a single teacher agent to suggest errors and fixes to the learners based on simulation results, achieving 80\% $pass@1$.

VerilogCoder \cite{ho2024verilogcoder} and MAGE \cite{zhao2024magemultiagentengineautomated} achieve some of the highest pass rates of agent-based approaches. VerilogCoder includes a planning agent, a plan verification assistant, a Verilog engineer agent, and Verilog verification assistant into its multi-agent frameworks. The task planner breaks down the natural language prompt into various subtask to implement, and the plan verification assistant reviews whether the tasks were implemented properly. This approach achieves a 94\% $pass@1$ rate on VerilogEval, far exceeding the roughly 63\% pass rate shown from state-of-the-art models in this work. However, VerilogCoder achieves this high pass rate by also having access to testbenches, simulation, and waveforms, so it can automatically debug a failing case until it passes. This is far more data and compute resources than what is given to models in single-turn, and individual models still have much room for improvement. Even more recently, MAGE leverages multiple agents, including a testbench agent, RTL agent, judge agent, and debug agent. Multiple RTL candidates are generated, ranked, and refined iteratively. MAGE's approach achieves 95\% pass rate on VerilogEval. 

Agentic approaches mimic the implement-debug cycle that human designers employ. Additionally, real-world design specifications are far from perfect, and often have bugs and ambiguities themselves. LLM-based agent flows will be essential in solving future real-world hardware design problems that move beyond trivial toy examples. Benchmarks must rise to meet these complexity needs and continue to aid in pushing the frontier, by presenting realistic design problems that are challenging for AI agents to solve. VerilogEval cannot meet this challenge, as the dataset is already solved for VerilogCoder and MAGE.

The construction of future benchmarks to address the needs demanded by agentic flows will not be trivial: realistic and complex design problems are needed, and they must not be overly descriptive but instead allow for reasoning and flexibility that a human designer would be afforded. This may include an assumed set of base knowledge on traditional digital design structures and best practices, but afford microarchitectural decision-making. An overly descriptive prompt that defines every wire or flip-flop will, at best, be a poor natural language proxy for RTL. Instead, benchmarks should include accurate behavioral specifications and designs goals, and allow an LLM agent to approach generating a solution within a reasonable decision space bounds.

\subsection{Related Hardware Design Tasks}

Hardware design is not limited to only RTL code generation, and many areas of hardware design would benefit from LLM enhancement within a tool flow. Even for RTL code generation, agentic approaches will demand specialized models or prompts for targeted tasks. This includes, but is not limited to, testbench creation \cite{zhang-llm4dv-arxiv2023, mi2024promptvleveragingllmpoweredmultiagent}, assertion generation \cite{orenesvera-autosva-dac2021}, documentation generation, debugging \cite{tsai-rtlfixer-arxiv2023, thakur-autochip-arxiv2023}, code review, analyzing consistency of specifications, correspondence of the testbench or RTL to test plan or specification, microarchitectural optimization, Q\&A, and many more \cite{fu-gpt4aigchip-iccad2023}. However, overall benchmark development for frontend hardware tasks beyond code generation is still early, and future benchmarks should strive to thoroughly and systematically evaluate many tasks.

There are fundamental differences between RTL and testbench code generation that must be addressed by benchmarks, models, and agents. For instance, Verilog testbench code is not constrained by the synthesizable subset of Verilog required for RTL. Moreover, verification code often adheres to coding conventions distinct from those used for RTL within an organization. Evaluation metrics must extend beyond mere syntactical and functional correctness. For verification code, key metrics include coverage of the RTL under test, whereas for RTL, the primary focus is on quality metrics such as power, performance, and area.

In essence, assessing LLMs and agents solely on their language capabilities is insufficient for hardware design. Tackling complex, high-level hardware design tasks necessitates breaking them down into smaller, manageable subtasks. This decomposition is particularly suited to agent-based approaches that leverage specialized LLMs for specific subtasks. Identifying and addressing weaknesses in models for these subtasks requires the development of additional benchmarks tailored to these categories.
\vspace{3mm}

\section{Conclusions}

Much improvement has been observed in large-language models for hardware code generation since VerilogEval v1 \cite{liu2023verilogeval} was released. Furthermore, the enhanced VerilogEval v2 benchmark proposed in this work provides a more robust framework for evaluating the performance of large-language models (LLMs) on digital hardware code generation tasks. Llama-3.1 405B and GPT-4o have both pushed the state-of-the-art as open and commercial models, while domain-specific models such as RTL-Coder 6.7B and DeepSeek Coder 6.7B have offered impressive pass rates for their parameter size.

When evaluated on 0-shot in-context learning for code completion, Llama3.1 405B outperforms GPT-4o in the equivalent of $pass@1$ on the VerilogEval-human benchmark. However, as in-context examples are added to the prompt, GPT-4o achieves parity with Llama3.1 405B, though Llama3.1 benefits from being an open model. Most models showed performance improvements with the addition of in-context learning examples. However, exceptions were observed, such as the degradation of Llama3 70B’s performance from 0-shot to 1-shot. Transitioning the benchmark task from code completion to specification-to-RTL generally yielded better results, as this approach grants models greater flexibility in defining interfaces. These findings highlight the critical role of task-specific tuning in enhancing code generation accuracy.

The improved benchmark infrastructure, including the new failure classification feature, provides more in-depth insights into the types of errors encountered by different models. For example, Llama3 70B frequently encounters \texttt{endmodule} missing errors during code completion, which careful prompt tuning or model alignment may be able to fix. The ability to classify and inspect failures on a problem-by-problem basis is critical for understanding and mitigating poor code generation, leading to more effective and targeted improvements in LLM performance for digital hardware code generation.

In the future, the research community would benefit from digital hardware benchmarks further expanded to include more tasks beyond RTL code generation representative of the digital hardware design flow. The enhanced VerilogEval v2 benchmark in this work is meant to be a step towards facilitating additional task support on top of a common set of design problems that allows for a more comprehensive assessment of model performance for hardware design. While the application of LLMs to hardware design is still in its infancy, models and generative AI techniques are quickly becoming capable of overcoming the code generation problems in the VerilogEval benchmark suite. It is imperative that as models and agents become more powerful, benchmark complexity increases to continue pushing technological development and sophistication in generative AI for hardware design.

\vspace{3mm}

\section*{Acknowledgment}
This paper would not have been possible without the generous help of NVIDIA Applied Deep Learning Research (ADLR), especially Teodor-Dumitru Ene, and the NVIDIA Inference Microservices (NIM) teams.

\bibliographystyle{ACM-Reference-Format}
\bibliography{ref}

\end{document}